\documentclass[11pt]{article}
\usepackage{mathrsfs}
\usepackage{amsmath}
\usepackage{amsfonts}
\usepackage{amssymb, amsmath, cite}
\usepackage{color}
\usepackage{graphicx}

\newcommand\be{\begin{equation}}
\newcommand\ee{\end{equation}}
\newcommand\ber{\begin{eqnarray}}
\newcommand\eer{\end{eqnarray}}
\newcommand\berr{\begin{eqnarray*}}
\newcommand\eerr{\end{eqnarray*}}
\newcommand\bea{\begin{eqnarray}}
\newcommand\eea{\end{eqnarray}}

\newcommand{\bfR}{{\Bbb R}}\newcommand{\ep}{\epsilon}

\newcommand{\x}{{\bf x}}

\newcommand{\dd}{\mbox{d}}\newcommand{\D}{\cal O}
\newcommand{\e}{\mbox{e}}
\newcommand{\pa}{\partial}

\newcommand{\nn}{\nonumber}

\newcommand\lb{\label}
\newcommand\eq{\eqref}

\title{Dyonically Charged Black Holes Arising in\\ Generalized Born--Infeld  Theory of Electromagnetism}

\author{
Yisong Yang\footnote{Email address: yisongyang@nyu.edu}\\Courant Institute of Mathematical Sciences\\ New York University\\New York, New York 10012, USA
}

\date{}

\begin{document}

\maketitle

\begin{abstract}
Black hole solutions to the Einstein equations coupled with the Born--Infeld electromagnetism associated with generalized nonlinear electrodynamics, carrying
both electric and magnetic charges, often referred to as dyonically charged black holes,  of finite energies, are constructed.  These solutions give rise to relegated curvature singularities at the center of the matter sources of the black holes and 
approach the classical Reissner--Nordstr\"{o}m black hole asymptotically.
\medskip

{\flushleft {Keywords:} Born--Infeld nonlinear electrodynamics, charged black holes, curvature singularities, regular
black holes, critical mass-energy condition, energy conditions
\medskip

\medskip

}
\end{abstract}

\section{Introduction}

It is well known that, in the charged Reissner--Nordstr\"{o}m black hole solution \cite{MTW,Wald,Carroll}, gravity and electromagnetism are treated rather differently in that
gravitational mass is finite but electromagnetic energy is infinite. In fact, this latter issue is associated with the structure of the Maxwell equations
in which a point charge carries an energy which is divergent at the spot where the charge resides, say, at the radial origin, $r=0$.  Specifically, using $(t,r,\theta,\phi)$ to denote
the coordinates of a spherically symmetric spacetime with the signature $(+---)$, then the Reissner--Nordstr\"{o}m metric assumes the form
\be\lb{1.1}
\dd s^2=\left(1-\frac{2GM}r+\frac{4\pi G Q^2}{r^2}\right)\dd t^2-\left(1-\frac{2GM}r+\frac{4\pi G Q^2}{r^2}\right)^{-1}\dd r^2-r^2\left(\dd\theta^2+\sin^2\theta\,\dd\phi^2\right),
\ee
where $Q>0$ is an effective charge parameter, $M>0$ the gravitational mass, $G$ Newton's gravitational constant, and the speed of light set to be unity. Hence it is clear that the charge
term does not contribute to the total mass but, on the other hand, aggravates or elevates the curvature singularity embodied in the Kretschmann invariant (say), $K$, from the type
$K\sim r^{-6}$ of the Schwarzschild black hole, to the type $K\sim r^{-8}$, in contrast.
At this juncture, we may recall that
the major motivation of Born and Infeld  \cite{B1,B2,B3,B4} for the introduction of their nonlinear
electromagnetic field theory
is to overcome the infinity problem associated with a  point charge source
in the original Maxwell theory in order to model the electron as a point charge. 
Thus, it would be natural to anticipate the black holes arising from coupling the Einstein equations with the Born--Infeld type nonlinear electrodynamics to enjoy improved or relegated
curvature singularity near the center of the black hole mass and charge, $r=0$. Along this line of development, regular black holes are obtained, in the frameworks of the
Bardeen black hole \cite{AG1,AG2,BV,Paula} and the Hayward black hole \cite{Hay,F,Kumar}. See \cite{JHOR} for a review on this and other related topics centered around the research on
the Born--Infeld theory suggested modified formalism of black hole constructions and cosmological applications. In a recent work \cite{Yang}, the author carried out a systematic study on the
relegation of the curvature singularity by the generalized Born--Infeld electrodynamics and established the general conclusion that a finite electromagnetic energy indeed boosts the
curvature regularity to the same level of that of the Schwarzschild black hole, $K\sim r^{-6}$, and that when the gravitational mass $M$ equals to the electromagnetic energy, $E$, 
a condition referred to as
the critical mass-energy condition, the curvature regularity may further be boosted. For example, for the rational-function model of Kruglov \cite{K7,K8,K9},  Ma \cite{Ma}
obtained regular magnetically charged black hole solutions, although their electrically charged counterparts are shown \cite{Yang}
to only enjoy a relegated curvature singularity, with $K\sim r^{-4}$. In fact, such a feature is already well observed in the Bardeen black hole \cite{AG1,AG2,BV,Paula}
and the Hayward black hole \cite{Hay,F,Kumar} situations where the regular solutions are magnetically charged but electrically charged ones are excluded, or vice versa, in view of
a correspondence relation called the $F$-$P$ duality, as a result of the electromagnetic asymmetry of the theory exhibited via the Minkowski spacetime signature. This setback further prevents
the construction of {\em both} electrically and magnetically charged black hole solutions, called dyonically charged black holes,  with a relegated curvature singularity in these models.
In fact, at the fundamental level, the main problem is that these models all fail to accommodate a finite-energy dyonically charged point source, an issue associated with the
limitation of the
first Born--Infeld model based solely upon a consideration of the action principle description of the free motion of a point mass in the context of relativistic mechanics (this model is also specifically referred to as the Born theory because of Born's own initial investigation \cite{B1,B2}). Fortunately, in \cite{B3,B4}, a geometric formalism based on an invariance principle consideration is also presented which  may be used to construct finite-energy dyonically charged
point sources \cite{Sch,W}. As a consequence, Demianski \cite{Dem} and Gibbons and Rasheed \cite{Gibbons-R} obtained finite-energy dyonically charged black holes in such a context.  Inspired by
the work of constructing regular magnetically charged black holes  \cite{AG1,AG2,BV,Paula,Hay,F,Kumar,JHOR,Ma} and 
charged black holes with relegated curvature singularities \cite{Yang} based on the generalized Born--Infeld nonlinear electrodynamics,  the goal of this study is to obtain finite-energy
{\em dyonically} charged black hole solutions to the Einstein equations coupled with the generalized Born--Infeld electromagnetism based on the invariance principle consideration \cite{B3,B4}.
We shall see that, unlike the Reissner--Nordstr\"{o}m charged black hole solution \eq{1.1}, these solutions all enjoy relegated curvature singularities with the same order as that of the Schwarzschild black hole at the center of the sources as a consequence of their finite electromagnetic energies and all approach the Reissner--Nordstr\"{o}m charged black hole solution \eq{1.1}
asymptotically near spatial infinity. Moreover, we shall see that there is also a critical mass-energy condition under which the curvature singularity may further be relegated beyond that of the Schwarzschild
black hole. Using the quasilocal energy formalism of Brown and York \cite{BY}, we shall also see that, like electric and magnetic charges of a Reissner--Nordstr\"{o}m black hole,
 the electromagnetic energy of a dyonically charged black hole in the generalized
Born--Infeld model does not contribute to the gravitational (ADM) mass \cite{MTW,Wald,Carroll,ADM1959,ADM,ADM1962} of the system.

The rest of the content of this work is outlined as follows. In Section 2, we review the Born--Infeld theory based on the invariance principle and formulate our generalized nonlinear
electrodynamics to be used to produce dyonically charged black holes of finite electromagnetic energies in subsequent sections.  In Section 3, we consider two concrete examples of dyonic point
charges of finite energies, one
based on the classical Born--Infeld theory \cite{B3,B4}, and another based on the exponential model developed by Hendi \cite{H1,H2}, with an additional feature that our models contain two
independent parameters so that we will see that the finite-energy property of the models may be turned on and off readily.
In Section 4, we obtain dyonically charge black hole solutions to the Einstein equations coupled with the generalized Born--Infeld nonlinear electrodynamics
under consideration and describe how to achieve
various levels of relegated curvature singularities in such a general setting. In particular,  we show that a critical mass-energy condition is present under which the curvature singularities may further be relegated
or even removed.
We also briefly describe how the Hawking radiation temperature \cite{FK,MW} and the Brown--York quasilocal energy \cite{BY} may be computed conveniently in this framework.
In Section 5, we use the explicit dyon particle constructions worked out in Section 3 to obtain the exact dyonically charge black hole solutions of the Einstein equations coupled with
the classical Born--Infeld model and the exponential model. In the latter situation and under the critical mass-energy condition, we show that the solution becomes regular only when the dyonic charge degenerates itself into a magnetic one as found in \cite{Yang}. In Section 6, we examine various energy conditions for the examples of the dyonically charged black holes obtained.
In Section 7, we summarize our results.

\section{Generalized Born--Infeld electromagnetism}

\setcounter{equation}{0}

 Let 
 $A_\mu$ be a real-valued gauge vector potential 
and 
\be\label{dyonick1.5}
F_{\mu\nu}=\pa_\mu A_\nu-\pa_\nu A_\mu
\ee
 the electromagnetic
field induced from $A_\mu$. Following the sign convention in \cite{Ryder,Huang2,Ybook}, we use
\be\label{dyonick1.6}
{\bf E}=(E^1, E^2, E^3),\quad E^i=-F^{0i};\quad {\bf B}=(B^1, B^2, B^3),\quad B^i=-\frac12\ep^{ijk}F_{jk},
\ee
to denote the electric and magnetic fields, respectively, where the standard Minkowski metric tensor $(\eta_{\mu\nu})$
or $(\eta^{\mu\nu})$ of the signature $(+---)$  is taken to lower or raise coordinate indices.
It will also be informative and convenient for calculation to rewrite \eqref{dyonick1.6} into its matrix form as
\be\label{dyonick1.8}
(F^{\mu\nu})=\left(\begin{array}{cccc}0&-E^1&-E^2&-E^3\\E^1&0&-B^3&B^2\\E^2&B^3&0&-B^1\\
E^3&-B^2&B^1&0\end{array}\right),
\ee
 such that 
the dual of $F^{\mu\nu}$ reads
\be\label{dyonick1.9}
{\tilde F}^{\mu\nu}=\frac12\ep^{\mu\nu\alpha\beta}F_{\alpha\beta}=\left(\begin{array}{cccc}0&-B^1&-B^2&-B^3\\B^1&0&
E^3&-E^2\\B^2&-E^3&0&E^1\\
B^3&E^2&-E^1&0\end{array}\right),
\ee
realizing via $F^{\mu\nu}\mapsto\tilde{F}^{\mu\nu}$ the usual electromagnetic duality ${\bf E}\mapsto{\bf B}, {\bf B}\mapsto -{\bf E}$.
So the Bianchi identity following from (\ref{dyonick1.5}) is
\be\label{dyonick1.10}
\pa_\mu {\tilde F}^{\mu\nu}=0.
\ee

In \cite{B3,B4}, Born and Infeld formulated a nonlinear geometric theory of electromagnetism purely based on
a consideration of the invariance principle imposed on the  action functional
\be\lb{dyonicxk32}
I=\int {\cal L}\,\dd x
\ee
over the Minkowski spacetime $(\bfR^{3,1},\eta_{\mu\nu})$ such that a simple and direct choice for the action density is to take
\be\lb{dyonicxk33}
{\cal L}=\gamma \sqrt{|\det(a_{\mu\nu})|},
\ee
where $\gamma$ is a coupling constant and $a_{\mu\nu}$ a certain covariant 2-tensor field which may be decomposed as a sum of a symmetric and  skewsymmetric
parts, say $s_{\mu\nu}$ and $f_{\mu\nu}$ for the moment. Motivated by simplicity again, we are prompted to choose
$s_{\mu\nu}=\eta_{\mu\nu}$ and $f_{\mu\nu}$ to be in proportion of the electromagnetic field  \eqref{dyonick1.5}. That is, we may set
\be\lb{dyonicxk34}
a_{\mu\nu}=\eta_{\mu\nu}+f_{\mu\nu},\quad f_{\mu\nu}=\kappa F_{\mu\nu}.
\ee
Besides, in order to maintain finiteness of the action, we need to subtract the `background' from \eqref{dyonicxk33} which pops up when
$f_{\mu\nu}$ vanishes such that we modify \eqref{dyonicxk33} into
\be\lb{dyonicxk35}
{\cal L}=\gamma \left(\sqrt{|\det(a_{\mu\nu})|}-\sqrt{|\det(\eta_{\mu\nu})|}\right)=\gamma \left(\sqrt{|\det(a_{\mu\nu})|}-1\right).
\ee
For convenience of reference,  from \eqref{dyonick1.8}, here we write down
\be\label{dyonicxk36}
(F_{\mu\nu})=\left(\begin{array}{cccc}0&E^1&E^2&E^3\\-E^1&0&-B^3&B^2\\-E^2&B^3&0&-B^1\\
-E^3&-B^2&B^1&0\end{array}\right),
\ee
since $F_{0i}=-F^{0i}, F_{ij}=F^{ij}$.  Thus, using the skewsymmetry of $f_{\mu\nu}$, \eqref{dyonicxk34}, and \eqref{dyonicxk36} , we have the expansion
\bea\lb{dyonicxk37}
\det(a_{\mu\nu})&=&\det(\eta_{\mu\nu}+f_{\mu\nu})\nn\\
&=&-1+(f_{01}f_{23}-f_{02}f_{13}+f_{03}f_{12})^2+(f_{01}^2+f_{02}^2+f_{03}^2-f_{12}^2-f_{13}^2-f_{23}^2)\nn\\
&=&-1+\kappa^2({\bf E}^2-{\bf B}^2)+\kappa^4 ({\bf E}\cdot{\bf B})^2.
\eea
Inserting \eqref{dyonicxk37} into \eqref{dyonicxk35}, we arrive at
\be\lb{dyonicxk38}
{\cal L}=-\gamma\left(1-\sqrt{1-\kappa^2({\bf E}^2-{\bf B}^2)-\kappa^4 ({\bf E}\cdot{\bf B})^2}\right),
\ee
where the choice of sign under the radical root is made in order to achieve consistency in the weak electromagnetic field limit.
Consequently, comparing \eqref{dyonicxk38} with the Maxwell action density 
\be\lb{dyonick1.3}
{\cal L}_{\mbox{\small M}}=-\frac14 F_{\mu\nu}F^{\mu\nu}=\frac12({\bf E}^2-{\bf B}^2), 
\ee
we are led to the recognition
\be\lb{dyonicxk39}
\gamma=-\frac1{\kappa^2},
\ee
such that \eqref{dyonick1.3} appears like a leading-order truncated theory model of \eqref{dyonicxk38} when $\kappa$ is small or the electromagnetic fields are weak.

On the other hand, from \eqref{dyonick1.9} and \eqref{dyonicxk36}, we have
\be\lb{dyonicxk41}
F_{\mu\nu}\tilde{F}^{\mu\nu}=-4\,{\bf E}\cdot{\bf B}.
\ee
Hence, by virtue of \eqref{dyonicxk38}--\eqref{dyonicxk41}, we obtain \cite{B3,B4}
\be\label{dyonic2.16}
{\cal L}=\frac1{\kappa^2}\left(1-\sqrt{1+\frac{\kappa^2}{2}F_{\mu\nu}F^{\mu\nu}-\frac{\kappa^4}{16}\left(F_{\mu\nu}\tilde{F}^{\mu\nu}\right)^2}\right).
\ee

For the purpose of our generalized formalism, we consider the action density
\be\lb{dyonic2.17}
{\cal L}=f(s), \quad s=-\frac14 F_{\mu\nu}F^{\mu\nu}+\frac{\kappa^2}{32}\left(F_{\mu\nu}\tilde{F}^{\mu\nu}\right)^2,
\ee
where the action density profile function $f$ is assumed to satisfy the condition
\be\lb{dyonic2.18}
f(0)=0,\quad f'(0)=1.
\ee
Hence the Born--Infeld theory \eq{dyonic2.16} gives
\be\lb{dyonic2.19}
f(s)=\frac1{\beta}\left(1-\sqrt{1-2\beta \,s}\right),
\ee
where and in the sequel we choose $\beta>0$ to be a coupling parameter independent of the parameter $\kappa$ for greater generality such that the limiting situation $\beta=\kappa^2$ returns the
model to the classical theory \eq{dyonic2.16} and reduces the computation considerably sometimes. The advantage of such a formalism is that \eq{dyonic2.17} allows the return to the 
generalized
model \cite{Yang}
\be\lb{dyonic2.17b}
{\cal L}=f(s), \quad s=-\frac14 F_{\mu\nu}F^{\mu\nu},
\ee
based on the classical Born model \cite{B1,B2} readily by setting $\kappa=0$ in \eq{dyonic2.17}.

Varying $A_\nu$ (say) in \eqref{dyonic2.17} leads to the associated Euler--Lagrange equations
\bea
\pa_\mu P^{\mu\nu}&=&0,\lb{dyonicxk43}\\
P^{\mu\nu}&=&f'(s)\left(F^{\mu\nu}-\frac{\kappa^2}{4}[F_{\alpha\beta}\tilde{F}^{\alpha\beta}]\tilde{F}^{\mu\nu}\right).\lb{dyonicxk44}
\eea

Corresponding to the  electric field  $\bf E$ and magnetic field $\bf B$ given in \eqref{dyonick1.6},  the usual
 electric displacement field $\bf D$ and  magnetic intensity field $\bf H$ are
\be\label{dyonick1.13}
{\bf D}=(D^1, D^2, D^3),\quad D^i=-P^{0i};\quad {\bf H}=(H^1, H^2, H^3),\quad H^i=-\frac12\ep^{ijk}P_{jk},
\ee
which may be suppressed into the  matrix form
\be\label{dyonick1.15}
(P^{\mu\nu})=\left(\begin{array}{cccc}0&-D^1&-D^2&-D^3\\D^1&0&-H^3&H^2\\D^2&H^3&0&-H^1\\
D^3&-H^2&H^1&0\end{array}\right).
\ee

Inserting \eq{dyonick1.9} into \eq{dyonick1.10} and \eq{dyonick1.15} into \eq{dyonicxk43}, respectively, we obtain
the  governing  equations of the following generalized Born--Infeld
 electromagnetic theory
\bea
\frac{\pa{\bf B}}{\pa t}+\nabla\times{\bf E}={\bf 0},\quad \nabla\cdot{\bf B}=0,\label{dyonick1.16}\\
-\frac{\pa{\bf D}}{\pa t}+\nabla\times {\bf H}={\bf 0},\quad\nabla\cdot{\bf D}=0,\label{dyonick1.17}
\eea
which take the same form of the vacuum Maxwell equations, except that, in view of the relation \eq{dyonicxk44}
 the electric displacement field $\bf D$ and magnetic intensity field $\bf H$ are now expressed as
\bea
{\bf D}&=&f'(s)\left({\bf E}+{\kappa^2}({\bf E}\cdot{\bf B})\,{\bf B}\right), \lb{dyonicxk48} \\
{\bf H}&=&f'(s)\left({\bf B}-{\kappa^2}({\bf E}\cdot{\bf B})\,{\bf E}\right),\lb{dyonicxk49}
\eea
in view of \eqref{dyonick1.8}, \eqref{dyonick1.9}, \eq{dyonicxk44}, and \eq{dyonick1.15}, which blend the electric  and magnetic fields, ${\bf E}$ and $\bf B$, and are the constitutive
equations in the current context of the generalized Born--Infeld electrodynamics.

From \eq{dyonic2.17}, we see that the associated energy-momentum tensor  is
\be\lb{dyonicxxbi98}
T_{\mu\nu}=-f'(s)\left(F_{\mu\alpha}\eta^{\alpha\beta}F_{\nu\beta}-\frac{\kappa^2}4(F_{\mu'\nu'}\tilde{F}^{\mu'\nu'})F_{\mu\alpha}\eta^{\alpha\beta}\tilde{F}_{\nu\beta}\right)-\eta_{\mu\nu}f(s).
\ee
Inserting \eq{dyonick1.8}, \eq{dyonick1.9}, \eq{dyonicxk36}, and \eq{dyonicxk41} into \eq{dyonicxxbi98}, we can write the energy density as
\bea
{\cal H}&=&T_{00}=
f'(s)\left({\bf E}^2+\kappa^2[{\bf E}\cdot{\bf B}]^2\right)
-f(s),\lb{dyonic2.29}\\
s&=&\frac12({\bf E}^2-{\bf B}^2)+\frac{\kappa^2}2({\bf E}\cdot{\bf B})^2,\lb{dyonic2.30}
\eea
which will be useful later in various calculations involving the energy of a dyonically charged source.

\section{Dyonically charged  point sources}
\setcounter{equation}{0}

As an illustration, we first consider the classical Born--Infeld model defined by \eq{dyonic2.19}.
With \eq{dyonic2.30} and some suitable manipulation of \eq{dyonicxk48}, we obtain
\bea
&&\left(1+\beta{\bf D}^2\right){\bf E}^2+\kappa^2\left(2+\beta{\bf D}^2+\kappa^2{\bf B}^2\right)({\bf E}
\cdot{\bf B})^2
={\bf D}^2\left(1+\beta{\bf B}^2\right),\lb{dyonicxxbi91}\\
&&\beta({\bf B}\cdot{\bf D})^2{\bf E}^2+\left(\left[1+\kappa^2{\bf B}^2\right]^2+\beta\kappa^2({\bf B}\cdot{\bf D})^2\right)({\bf E}\cdot{\bf B})^2
=({\bf B}\cdot{\bf D})^2\left(1+\beta{\bf B}^2\right),\lb{dyonicxxbi92}
\eea
with the solution
\bea
{\bf E}^2&=&\frac{(\beta{\bf B}^2+1)({\bf D}^2+\kappa^2[2+\kappa^2{\bf B}^2][{\bf B}\times{\bf D}]^2)}{(\kappa^2{\bf B}^2+1)(1+\kappa^2{\bf B}^2+\beta{\bf D}^2+\beta\kappa^2[{\bf B}\times{\bf D}]^2)},\lb{dyonicxxe}\\
({\bf E}\cdot{\bf B})^2&=&\frac{(\beta {\bf B}^2+1)({\bf B}\cdot{\bf D})^2}{(\kappa^2{\bf B}^2+1)(1+\kappa^2{\bf B}^2+\beta{\bf D}^2+\beta\kappa^2[{\bf B}\times{\bf D}]^2)},\lb{dyonicxxeb}
\eea
where we have used the identity 
\be
({\bf B}\times{\bf D})^2={\bf D}^2{\bf B}^2-({\bf B}\cdot{\bf D})^2.
\ee

We now consider a dyonic point particle with electric charge  $q>0$ and magnetic charge $g>0$ resting at the origin of the space such that the induced
electric displacement field and magnetic field are given by
\be\lb{dyonic3.6}
\nabla\cdot{\bf D}=4\pi q\delta({\bf x}),\quad \nabla\cdot{\bf B}=4\pi g\delta({\bf x}).
\ee
Thus the nontrivial radial components of ${\bf D}$ and $\bf B$ are obtained to be
\be\lb{dyonic3.7}
D^r=\frac q{r^2},\quad B^r=\frac g{r^2},\quad r=|\x|>0,
\ee
respectively.
Inserting \eq{dyonic3.7} into \eq{dyonicxxe} and \eq{dyonicxxeb},  we have
\be\lb{dyonic3.8}
{\bf E}^2=\frac{(\beta g^2+r^4)q^2}{(\kappa^2 g^2+r^4)(\beta q^2+\kappa^2 g^2+r^4)},\quad ({\bf E}\cdot{\bf B})^2=\frac{(\beta g^2+r^4)q^2 g^2}{r^4(\kappa^2 g^2+r^4)(\beta q^2+\kappa^2 g^2+r^4)},
\ee
which in particular gives us the nontrivial radial component of $\bf E$ to be 
\be\lb{dyonic3.9}
E^r=q\sqrt{\frac{\beta g^2+r^4}{(\kappa^2 g^2+r^4)(\beta q^2+\kappa^2 g^2+r^4)}},
\ee
which remains finite at $r=0$, although $E^r$ still obeys the
Coulomb law, $E^r\sim \frac{q}{r^2}$, for $r$ near infinity. Besides, the free electric charge density reads
\bea\lb{dyonic3.10}
\rho^e_{\mbox{\tiny free}}&=&\frac1{4\pi}\nabla\cdot{\bf E}=\frac1{4\pi r^2}\frac{\dd}{\dd r}(r^2 E^r)\nn\\
&=&\frac q{2\pi r}\frac{\left(\beta \kappa^2 g^4(\beta q^2+\kappa^2 g^2)+2\kappa^2 g^2(\beta q^2+\kappa^2 g^2)r^4+(2\kappa^2 g^2+\beta[q^2-g^2])r^8\right)}{ (\beta g^2+r^4)^{\frac12}(\kappa^2 g^2+r^4)^{\frac32}(\beta q^2+\kappa^2 g^2+r^4)^{\frac32}},
\eea
such that
\be\lb{dyonicrho-e}
\rho^e_{\mbox{\tiny free}}=\frac q{2\pi \kappa r}\sqrt{\frac{\beta}{\beta q^2+\kappa^2 g^2}},\quad r\ll1;\quad \rho^e_{\mbox{\tiny free}}=\frac q{2\pi r^7}\left(2\kappa^2 g^2
+\beta[q^2-g^2]\right),\quad r\gg1,
\ee
which clearly describe how  electricity and magnetism mix and interact with each other through the induced free electric charge density, both locally and asymptotically.
 Furthermore, from \eq{dyonic3.9} and \eq{dyonic3.10}, the free electric charge contained in $\{|\x|\leq r\}$ is seen to be
\bea
q_{\mbox{\tiny free}}(r)&=&\int_{|\x|\leq r}\rho^e_{\mbox{\tiny free}}\,\dd\x\nn\\
&=&(r^2 E^r)_{r=0}^{r=r}=qr^2\sqrt{\frac{\beta g^2+r^4}{(\kappa^2 g^2+r^4)(\beta q^2+\kappa^2 g^2+r^4)}},
\eea
such that the total free electric charge coincides with the prescribed one, $q_{\mbox{\tiny free}}=q_{\mbox{\tiny free}}(\infty)=q$.

On the other hand, using \eq{dyonic3.7}, \eq{dyonic3.8}, and \eq{dyonic3.9} in \eq{dyonicxk49} with \eq{dyonic2.19} and \eq{dyonic2.30}, we get the nontrivial radial component of
the magnetic intensity field
\be\lb{dyonicHr}
H^r=\frac{g\left(\kappa^4 g^4+(2\kappa^2 g^2 +q^2[\beta-\kappa^2])r^4+r^8\right)}{(\beta g^2+r^4)^{\frac12}(\beta q^2+\kappa^2 g^2+r^4)^{\frac12}(\kappa^2 g^2+r^4)^{\frac32}},
\ee
which is parallel to the expression for the electric field stated in \eq{dyonic3.9} such that $H^r$ remains finite at $r=0$, and $H^r\sim \frac g{r^2}$ for $r\gg1$, as anticipated for a magnetic monopole. Using \eq{dyonicHr}, we may compute the associated free magnetic charge density to obtain
\be
\rho^m_{\mbox{\tiny free}}=\frac{\kappa g}{2\pi r\sqrt{\beta(\beta q^2+\kappa^2 g^2)}},\quad r\ll1; \quad \rho^m_{\mbox{\tiny free}}=\frac g{2\pi r^7}\left(2\kappa^2 q^2+\beta[g^2-q^2]\right),\quad r\gg1,
\ee
similar to \eq{dyonicrho-e}.
Besides, the total free magnetic charge
coincides with the prescribed magnetic charge as well,  $g_{\mbox{\tiny free}}=g$.

The expressions \eq{dyonic3.9} and \eq{dyonicHr} give us an exact full-parameter dyonic point source solution for the generalized model \eq{dyonic2.19}. This solution,
which appears to be new, exhibits the roles played by all the free parameters involved clearly.

We now consider the energy carried by such a dyonic point charge. Inserting \eq{dyonic2.19} and \eq{dyonic2.30} into \eq{dyonic2.29}, we have
\be
{\cal H}=\frac1\beta\left(\frac1{\cal R}-1\right)
+\frac{{\bf B}^2}{\cal R},\quad
{\cal R}=\sqrt{1-{\beta}({\bf E}^2-{\bf B}^2+{\kappa^2}[{\bf E}\cdot{\bf B}]^2)}.\lb{dyonic3.13}
\ee
Thus, substituting \eq{dyonicxxe} and \eq{dyonicxxeb} into \eq{dyonic3.13}, we arrive at
\be\lb{dyonicxxbih}
{\cal H}=\frac1{\beta}\left(\left[\frac{(1+{\beta{\bf B}}^2)(1+\beta({\bf D}^2+\kappa^2[{\bf D}\times{\bf B}]^2)+\kappa^2{\bf B}^2)}{1+\kappa^2{\bf B}^2}\right]^{\frac12}-1\right),
\ee
which is all given in terms of prescribable quantities.
Inserting \eq{dyonic3.7} into \eq{dyonicxxbih}, we have
\bea
{\cal H}&=&\frac1{\beta}\left(\sqrt{\frac{r^4+\beta g^2}{r^4+\kappa^2 g^2}}\sqrt{1+\frac{\beta q^2+\kappa^2 g^2}{r^4}}-1\right)\nn\\
&=&\frac{(\beta q^2+\kappa^2 g^2)g^2+(q^2+g^2)r^4}{ r^2\sqrt{r^4+\kappa^2 g^2}(\sqrt{r^4+\beta g^2}\sqrt{r^4+\beta q^2+\kappa^2 g^2}+r^2\sqrt{r^4+\kappa^2g^2})},\lb{dyonic3.15}
\eea
which enables the determination of the finite total energy $E$ of the dyonically charged point particle by the usual expression
\be\lb{dyonic3.16}
\frac{E}{4\pi}=\int_0^\infty r^2{\cal H}\,\dd r.
\ee

First, we note that, when $\kappa=0$, \eq{dyonic3.15} becomes
\be
{\cal H}_{\kappa=0}=\frac{\beta q^2 g^2+(q^2+g^2) r^4}{r^4(\sqrt{r^4+\beta q^2}\sqrt{r^4+\beta g^2}+r^4)}.
\ee
Hence, in the dyonic point charge situation, $q,g>0$, the energy \eq{dyonic3.16} diverges, although in the pure electric or magnetic point charge situation with $q>0,g=0$ or $q=0,g>0$, respectively,  the energy 
converges. In other words, the coupling parameter $\kappa$ plays the role of an energy regulator for a dyonic point charge, so that, when it is `switched off' by setting $\kappa=0$, the energy
blows up.

In contrast, when $\kappa>0$ but the point charge is either purely electric or magnetic, with $q>0, g=0$ or $q=0,g>0$, respectively, we see that the energy always converges.

Next, in the general dyonic situation with $q,g,\kappa>0$, we note that it is impossible to obtain the exact value for the energy $E$ given in \eq{dyonic3.16}.  However, in order to estimate $E$, we 
observe that the quantity
\be
\eta(r)=\sqrt{\frac{r^4+\beta g^2}{r^4+\kappa^2 g^2}}
\ee
is nondecreasing or nonincreasing in $r\geq 0$ according to whether $\beta\leq\kappa^2$ or $\beta\geq\kappa^2$. Consequently, we have the general bounds
\be
\min\left\{1,\frac{\sqrt{\beta}}\kappa\right\}\leq     \eta(r)     \leq\max\left\{1,\frac{\sqrt{\beta}}\kappa\right\},\quad r\geq0.
\ee
In particular, we have
\be
\eta(r)\leq 1,\quad \beta\leq\kappa^2; \quad \eta(r)\geq 1,\quad\beta\geq\kappa^2.
\ee
Applying these bounds to \eq{dyonic3.16}, we get the quantities
\be
E\leq E_0,\quad \beta\leq \kappa^2;\quad E\geq E_0,\quad \beta\geq\kappa^2,
\ee
where
\bea\lb{dyonic3.23a}
\frac{E_0}{4\pi}&=&\frac1\beta \int_0^\infty\left(\sqrt{1+\frac{\beta q^2+\kappa^2 g^2}{r^4}}-1\right) r^2\,\dd r\nn\\
&=&\frac{a^3}{\beta}\int_0^\infty\frac{\dd x}{\sqrt{x^4+1}+x^2}\nn\\
&=&\frac{\pi^{\frac32}}{3\left(\Gamma\left(\frac34\right)\right)^2}\,\frac{Q^2}a,
\eea
where we set
\be\lb{dyonic3.17}
Q=\left(q^2+\left[\frac{\kappa^2}{\beta}\right]g^2\right)^{\frac12},\quad a=(\beta q^2+\kappa^2 g^2)^{\frac14},
\ee
which are to be viewed as some estimates of the {\em combined dyonic charge}\index{combined dyonic charge} taking account of both electric and magnetic charges, $q$
and $g$, jointly, residing at the same spot, and  of the `radius' of the
dyonically charged particle, respectively. Note that, when $\beta=\kappa^2$, we recover the classical result.

It is interesting to see that  \eq{dyonic3.23a} is of the same form of that for an electric point charge, as in the work of Born and Infeld \cite{B4}. In other words, in terms of the combined charge defined in \eq{dyonic3.17}, a dyonic point charge behaves like an electric point charge, energetically. Furthermore, the description of a dyonic point charge is symmetric with respect
to the interchange of the roles of the electric and magnetic charges in the classical-model limit $\beta=\kappa^2$.
Besides, the second expression in \eq{dyonic3.17} indicates that a dyonic point charge possesses a larger radius as a result of the
need to accommodate a greater
joint charge as given in the first expression in \eq{dyonic3.17}.

Although \eq{dyonic3.15} appears complicated, it has the following simple expansion
\be\lb{dyonic3.25+}
{\cal H}=\frac{q^2+g^2}{2r^4}-\frac{\left(\beta(q^2-g^2)^2 +4\kappa^2 q^2 g^2\right)}{8r^8}+\mbox{O}(r^{-12}),\quad r\gg1,
\ee
extending that given by the Maxwell theory,  ${\cal H}=\frac{q^2+g^2}{2r^4}$ for $r>0$, for a dyonic point charge.

As another illustration, we next consider the exponential model \cite{H1,H2}  given by
\be\lb{dyonic3.18}
f(s)=\frac1{\beta}\left(\e^{\beta s}-1\right),
\ee
where $\beta>0$ is again a parameter. With this and \eq{dyonic2.30}, the equations \eq{dyonicxk48} and \eq{dyonicxk49} become
\bea{\bf D}&=&\e^{\frac\beta2\left({\bf E}^2-{\bf B}^2+\kappa^2[{\bf E}\cdot{\bf B}]^2\right)}\left({\bf E}+\kappa^2[{\bf E}\cdot{\bf B}]{\bf B}\right),\lb{dyonic3.19}\\
{\bf H}&=&\e^{\frac\beta2\left({\bf E}^2-{\bf B}^2+\kappa^2[{\bf E}\cdot{\bf B}]^2\right)}\left({\bf B}-\kappa^2[{\bf E}\cdot{\bf B}]{\bf E}\right).\lb{dyonic3.20}
\eea
From \eq{dyonic3.19}, we have
\bea
{\bf D}^2 \e^{\beta{\bf B}^2}&=&\e^{\beta({\bf E}^2+\kappa^2[{\bf E}\cdot{\bf B}]^2)}\left({\bf E}^2+\kappa^2[2+\kappa^2{\bf B}^2][{\bf E}\cdot{\bf B}]^2\right),\lb{dyonic3.21}\\
({\bf B}\cdot{\bf D})^2 \e^{\beta{\bf B}^2}&=&\e^{\beta({\bf E}^2+\kappa^2[{\bf E}\cdot{\bf B}]^2)}\left(1+\kappa^2{\bf B}^2\right)^2({\bf E}\cdot{\bf B})^2.\lb{dyonic3.22}
\eea
Note that the left-hand sides of these two equations are prescribable in view of \eq{dyonic3.7} so that some manipulation of these equations renders the equation
\be
\e^W W=\delta,\quad W=\beta\left({\bf E}^2+\kappa^2[{\bf E}\cdot{\bf B}]^2\right),\quad\delta=\beta\e^{\beta{\bf B}^2}\left({\bf D}^2-\frac{\kappa^2({\bf B}\cdot{\bf D})^2}{1+
\kappa^2{\bf B}^2}\right),\lb{dyonic3.23}
\ee
which may be solved with the Lambert $W$ function \cite{Chow,CG}
\bea
\beta\left({\bf E}^2+\kappa^2[{\bf E}\cdot{\bf B}]^2\right)&=&W(\delta)=W\left(\beta\e^{\beta{\bf B}^2}\left[{\bf D}^2-\frac{\kappa^2({\bf B}\cdot{\bf D})^2}{1+
\kappa^2{\bf B}^2}\right]\right)\nn\\
&=&W\left(\beta\e^{\beta{\bf B}^2}\left[\frac{{\bf D}^2+\kappa^2({\bf B}\times{\bf D})^2}{1+
\kappa^2{\bf B}^2}\right]\right),\lb{dyonic3.24}
\eea
where $W(x)$ is well defined and analytic for $x>-\frac1\e$ with the Taylor expansion
\be\lb{dyonic3.25}
W(x)=\sum_{n=1}^\infty \frac{(-n)^{n-1}}{n!} x^n,
\ee
about $x=0$, and the asymptotic expansion
\be\lb{dyonic3.26}
W(x)=\ln x-\ln\ln x+\frac{\ln\ln x}{\ln x}+\cdots,\quad x>3,
\ee
which are sufficient for our purposes. Furthermore, by \eq{dyonic3.22}, we get
\be\lb{dyonic3.27}
({\bf E}\cdot{\bf B})^2=\frac{({\bf B}\cdot{\bf D})^2}{(1+\kappa^2{\bf B}^2)^2}{ \e^{\beta{\bf B}^2-W(\delta)}}.
\ee
Combining \eq{dyonic3.24} and \eq{dyonic3.27}, we arrive at
\be\lb{dyonic3.28}
{\bf E}^2=\frac{W(\delta)}\beta-\frac{\kappa^2({\bf B}\cdot{\bf D})^2}{(1+\kappa^2{\bf B}^2)^2}{ \e^{\beta{\bf B}^2-W(\delta)}},
\ee
where $W(\delta)$ is as stated in \eq{dyonic3.24}. This expression appears rather complicated. However, in the radially symmetric dyonically charged point particle situation given by \eq{dyonic3.7}, 
the quantity $\delta$ in \eq{dyonic3.23} or \eq{dyonic3.24} simplifies into
\be\lb{dyonic3.29}
\delta=\frac{\beta q^2}{r^4+\kappa^2 g^2}\, \e^{\frac{\beta g^2}{r^4}}.
\ee
In view of \eq{dyonic3.25} and \eq{dyonic3.29}, we have
\be\lb{dyonic3.30}
W(\delta)=\frac{\beta q^2}{r^4}\left(1+\frac1{r^4}\left[(\beta -\kappa^2)g^2-{\beta q^2}\right]\right)+\mbox{O}(r^{-12}),\quad r\gg1,
\ee
such that the nontrivial radial component $E^r$ of the electric field $\bf E$ reads 
\be\lb{dyonic3.31}
E^r=\frac{q}{r^2}+\frac q{r^6}\left(\frac\beta2(g^2-q^2)-\kappa^2 g^2\right)+\mbox{O}(r^{-10}),\quad r\gg1,
\ee
asymptotically. In other words, the electric field of the dyonic point charge appears like that of an electric Coulomb charge asymptotically. On the other hand, for $\kappa>0$,  in view of \eq{dyonic3.26} and \eq{dyonic3.29}, we have
\be\lb{dyonic3.40}
W(\delta)=\frac{\beta g^2}{r^4}+\ln r^4+\ln\left(\frac{q^2}{\kappa^2 g^4}\right)-\frac1{g^2}\left(\frac1{\kappa^2}+\frac{1}{\beta}\ln\left[\frac{\beta q^2}{\kappa^2 g^2}\right]\right)r^4+\mbox{o}(r^8),\quad r\ll1,
\ee
in leading-order approximation, which results in the behavior
\be\lb{dyonic3.33}
E^r=\frac{g}{r^2}\sqrt{1-\frac{\kappa^4 g^4}{(r^4+\kappa^2 g^2)^2}},\quad r\ll1.
\ee
In other words, locally, the electric field of the dyonic point charge depends only on its magnetic charge $g$. As a consequence of \eq{dyonic3.31} and \eq{dyonic3.33}, we see that the total
free electric charge of the particle is
\be
q_{\mbox{\tiny free}}=\left(r^2 E^r\right)_{r=0}^{r=\infty}=q,
\ee
which coincides with the prescribed electric charge, rather surprisingly, because the electric field of a dyonic point charge locally depends only on its magnetic charge as seen in \eq{dyonic3.33}.

The limiting situation $\kappa=0$ calls for a separate treatment. Now \eq{dyonic3.29} becomes
\be
\delta=\frac{\beta q^2}{r^4}\e^{\frac{\beta g^2}{r^4}},
\ee
which leads to
\be\lb{dyonicW45}
W(\delta)=\frac{\beta g^2}{r^4}+\ln\frac{q^2}{g^2}-\frac{r^4}{\beta g^2}\ln\left(\frac{\beta q^2}{r^4}\right)+\mbox{O}(r^8),\quad r\ll1.
\ee
Hence \eq{dyonic3.28} gives rise to the simple result
\be\lb{dyonicE46}
E^r=\frac{g}{r^2},\quad r\ll1,
\ee
in leading order approximation, indicating that the electric field of that of a dyonic point particle behaves like that of a magnetic point charge locally and the associated total free electric charge 
is then given by
\be\lb{q}
q_{\mbox{\tiny free}}=(r^2 E^r)_{r=0}^{r=\infty}=q-g,
\ee
instead, which mixes the electric and magnetic charges. Thus, in this regard, the parameter $\kappa$ is seen to play the role of switching off such a mixing effect.

Furthermore, in view of \eq{dyonic3.20} and \eq{dyonic3.31}, we see that the nontrivial radial component of the magnetic intensity field $\bf H$ is
\be\lb{dyonic3.35}
H^r=\frac g{r^2}+\frac{g}{r^6}\left(\frac\beta2(q^2-g^2)-\kappa^2 q^2\right)+\mbox{O}(r^{-10}),\quad r\gg1,
\ee
in leading-order approximation, which indicates that this field asymptotically behaves like that of a magnetic point charge. Moreover, there is a perfect symmetry $E^r\leftrightarrows H^r$,
$q\leftrightarrows g$ between the two formulas,  \eq{dyonic3.31} and \eq{dyonic3.35}, for $E^r$ and $H^r$, respectively. 

On the other hand, when $\kappa>0$,  inserting \eq{dyonic3.40} and \eq{dyonic3.33} into  \eq{dyonic3.20}, we have, up to leading orders, the expression
\bea\lb{dyonic3.45}
H^r&=&\frac{q}{\kappa g}\left(1-\frac{\kappa^2 g^2}{r^4}\left[1-\frac{\kappa^4 g^4}{(r^4+\kappa^2 g^2)^2}\right]\right)\nn\\
&=&\frac{q}{\kappa g}\left(-1+\frac{3r^4}{\kappa^2 g^2}\right)+\mbox{O}(r^8),\quad r\ll1.
\eea
In particular, $H^r$ stays finite near the origin and the associated total free magnetic charge is calculated to be
\be\lb{g}
g_{\mbox{\tiny free}}=(r^2 H^r)_{r=0}^{r=\infty}=g,
\ee
which coincides with the prescribed point magnetic charge, by using \eq{dyonic3.35} and \eq{dyonic3.45} as before.  Interestingly, unlike \eq{dyonic3.33} for $E^r$, the expression 
\eq{dyonic3.45} involves both electric and magnetic charges, although the final, global, result \eq{g} is solely magnetic.

We again consider $\kappa=0$ separately. Inserting \eq{dyonicW45} and \eq{dyonicE46} into \eq{dyonic3.20}, we have the following leading-order approximation,
\be\lb{dyonicH51}
H^r=\frac q{r^2},\quad r\ll1,
\ee
which is similar to \eq{dyonicE46} such that the induced magnetic intensity field resembles the Coulomb field of an electric point charge near the origin and that 
the associated total free magnetic charge reads
\be
g_{\mbox{\tiny free}}=(r^2 H^r)_{r=0}^{r=\infty}=g-q,
\ee
in view of \eq{dyonic3.35} (with $\kappa=0$) and \eq{dyonicH51}. Thus it is seen as another mixed charge analogous to the free electric charge \eq{q}.

To compute the energy of a dyonic point charge in the exponential model, we insert \eq{dyonic3.18} into \eq{dyonic2.29}--\eq{dyonic2.30} to get the Hamiltonian density
\be\lb{dyonicH-exp}
{\cal H}=\frac1\beta \e^{-\frac\beta2{\bf B}^2+\frac12 W(\delta)}\,W(\delta)-\frac1\beta\left(\e^{-\frac\beta2{\bf B}^2+\frac12 W(\delta)}-1\right).
\ee
Consequently,  in view of \eq{dyonic3.30}, we have
\be\lb{dyonicH-exp2}
{\cal H}=\frac{q^2+g^2}{2r^4}-\frac{\left(\beta[q^2-g^2]^2+4\kappa^2 q^2 g^2\right)}{8r^8}+\mbox{O}(r^{-12}),\quad r\gg1,
\ee
which in leading order is again identical to that in the Maxwell electromagnetism theory, and incidentally coincides with \eq{dyonic3.25+}. Moreover, we have
\be\lb{dyonicH-exp3}
{\cal H}=\frac{q}{\kappa r^2}+\frac1\beta-\frac{qr^2}{2\beta \kappa^3 g^2}\left(\beta+\kappa^2\left[2+\ln\left(\frac{\beta\kappa^2 g^6}{q^2 r^6}\right)\right]\right)+\mbox{O}(r^6),\quad r\ll1,\quad\kappa>0;
\ee
\be
{\cal H}=\frac{qg}{r^4},\quad r\ll1,\quad \kappa=0.
\ee
As a consequence of these results, we conclude that  there is  energy convergence when $\kappa>0$ but divergence when $\kappa=0$ and $q,g>0$. In other words, the coupling parameter $\kappa$ plays the role of an energy regulator for a dyonic point charge as before as that for the classical Born--Infeld theory \eq{dyonic2.19}.

This exact dyonic point source solution and its refined parameter-dependent properties unveiled for the exponential model \eq{dyonic3.18} appear to be new.

\section{Dyonically charged black hole solutions in general setting}
\setcounter{equation}{0}

In this section, we consider dyonically charged black hole solutions generated from the Einstein equations with a vanishing cosmological constant
coupled with the Born--Infeld equations associated with a generalized action density,  \eq{dyonic2.17}, in which the Minkowski metric $\eta_{\mu\nu}$ is replaced
by a gravitational one, $g_{\mu\nu}$,  such that, with $g=\det(g_{\mu\nu})$,  which should not be confused with the magnetic charge under the same notation and should be clear in
the context, we have
\be\lb{dyonic4.1}
 F_{\mu\nu}=g_{\mu\alpha}g_{\nu\beta}F^{\alpha\beta},\quad \tilde{F}^{\mu\nu}=\frac12\varepsilon^{\mu\nu\alpha\beta}F_{\alpha\beta},\quad \varepsilon^{0123}=\frac1{\sqrt{-g}}.
\ee
The curved-spacetime Born--Infeld electromagnetic field equations of \eq{dyonic2.17} with \eq{dyonic4.1} now read
\bea
\frac1{\sqrt{-g}}\pa_\mu\left(\sqrt{-g}P^{\mu\nu}\right)&=&0,\lb{dyonic4.2}\\
P^{\mu\nu}&=&f'(s)\left(F^{\mu\nu}-\frac{\kappa^2}4[F_{\alpha\beta}\tilde{F}^{\alpha\beta}]\tilde{F}^{\mu\nu}\right),\lb{dyonic4.3}
\eea
which update \eq{dyonicxk43}--\eq{dyonicxk44}.  Likewise, the energy-momentum tensor $T_{\mu\nu}$ is now given by
\bea\lb{dyonic4.4}
T_{\mu\nu}&=&2\frac{\pa{\cal L}}{\pa g^{\mu\nu}}-g_{\mu\nu}{\cal L}\nn\\
&=&-f'(s)\left(F_{\mu\alpha}g^{\alpha\beta}F_{\nu\beta}-\frac{\kappa^2}4(F_{\mu'\nu'}\tilde{F}^{\mu'\nu'})F_{\mu\alpha}g^{\alpha\beta}\tilde{F}_{\nu\beta}\right)-g_{\mu\nu}f(s).
\eea
Thus, using
$T=g^{\mu\nu}T_{\mu\nu}$ to denote the trace of $T_{\mu\nu}$,  the Einstein equations are
\be\lb{dyonic4.5}
R_{\mu\nu}=-8\pi G\left(T_{\mu\nu}-\frac12 g_{\mu\nu}T\right),
\ee
expressed in terms of the Ricci tensor $R_{\mu\nu}$ associated with the metric $g_{\mu\nu}$.

We are interested in dyonically charged black hole solutions to the coupled equations \eq{dyonic4.2}--\eq{dyonic4.5} such that the spacetime line element in the ordered
spherical coordinates $(x^\mu)=(t,r,\theta,\phi)$ assumes the Schwarzschild form
\be\lb{dyonic4.6}
\dd s^2=g_{\mu\nu}\dd x^\mu\dd x^\nu=A(r)\dd t^2-\frac{\dd r^2}{A(r)}-r^2\left(\dd\theta^2+\sin^2\theta\,\dd\phi^2\right).
\ee
With \eq{dyonic4.6}, the nontrivial components of the Ricci tensor are
\be\lb{dyonic4.7}
R_{00}=-\frac{A}{2r^2}(r^2 A')',\quad
R_{11}=\frac{1}{2r^2A}(r^2 A')',\quad
R_{22}=(rA)'-1,\quad
R_{33}=\sin^2\theta\, R_{22},
\ee
where and in the sequel the prime $'$ denotes differentiation with respect to the radial variable $r$ which should not be confused with that for $f(s)$
with respect to $s$ in \eq{dyonic4.3}--\eq{dyonic4.4} (say) and should cause no confusion in different contexts.

On the other hand, for the electromagnetic sector,  set
\be\lb{dyonic4.8}
(P^{\mu\nu})=\left(\begin{array}{cccc}0&-D^r&0&0\\D^r&0&0&0\\0&0&0&-H^r\\0&0&H^r&0\end{array}\right),
\ee
where $D^r$ and $H^r$ are the nontrivial radial components of the electric displacement and magnetic intensity fields, respectively, to be determined. Thus, by consistency in \eq{dyonic4.3}, we have
\be\lb{dyonic4.9}
(F^{\mu\nu})=\left(\begin{array}{cccc}0&-E^r&0&0\\E^r&0&0&0\\0&0&0&-B^r\\0&0&B^r&0\end{array}\right),
\ee
where the radial components of the electric and magnetic fields, $E^r$ and $B^r$, are to be determined.
Hence, with \eq{dyonic4.9} and \eq{dyonic4.1}, 
 we have
\bea
({F}_{\mu\nu})&=&\left(\begin{array}{cccc}0&E^r&0&0\\-E^r&0&0&0\\0&0&0&-r^4\sin^2\theta\, B^r\\0&0&r^4\sin^2\theta\, B^r&0\end{array}\right),\lb{dyonic4.10}\\
(\tilde{F}^{\mu\nu})&=&\left(\begin{array}{cccc}0&-r^2\sin\theta B^r&0&0\\r^2\sin\theta B^r&0&0&0\\0&0&0&\frac1{r^2\sin\theta} E^r\\0&0&-\frac1{r^2\sin\theta} E^r &0\end{array}\right).\lb{dyonic4.11}
\eea
Consequently we have
\bea
F_{\mu\nu}F^{\mu\nu}&=&-2(E^r)^2 +2r^4\sin^2\theta\, (B^r)^2,\lb{dyonic4.12}\\
F_{\mu\nu}\tilde{F}^{\mu\nu}&=& -4 r^2 \sin\theta\, (E^r B^r).\lb{dyonic4.13}
\eea
As a consequence, we get
\be\lb{dyonic4.14}
s=\frac12\left([E^r]^2-r^4\sin^2\theta [B^r]^2\right)+\frac{\kappa^2}2 r^4\sin^2\theta (E^r B^r)^2.
\ee
Besides, by \eq{dyonic4.3} and \eq{dyonic4.8}, we have
\be\lb{dyonic4.15}
D^r=f'(s)\left(E^r+\kappa^2 r^4\sin^2 \theta \, E^r [B^r]^2\right),\quad H^r=f'(s)\left(B^r-\kappa^2 [E^r]^2 B^r\right).
\ee
Moreover,  with \eq{dyonic4.6}, we have $\sqrt{-g}=r^2\sin \theta$ so that a combination of \eq{dyonic4.2} and \eq{dyonic4.8} yields the relations
\be\lb{dyonic4.16}
\pa_r (r^2\sin\theta D^r)=0,\quad
\pa_\theta (r^2\sin\theta H^r)=0,
\ee
resulting in 
\be\lb{dyonic4.17}
D^r=\frac q{r^2},
\ee
where $q$ is an integration constant which gives rise to an electric charge as before, which is taken to be positive for convenience, and
\be\lb{dyonic4.18}
H^r=\frac{H(r)}{\sin\theta},
\ee
where $H(r)$ is a function of the radial variable $r$ to be determined.  Thus, in view of consistency in \eq{dyonic4.15} with \eq{dyonic4.17} and \eq{dyonic4.18}, we arrive at
\be\lb{dyonic4.19}
B^r=\frac{B(r)}{\sin\theta},
\ee
which resembles \eq{dyonic4.18}, where the function $B(r)$ is to be determined. Hence \eq{dyonic4.12}--\eq{dyonic4.14} are simplified into
\bea
F_{\mu\nu}F^{\mu\nu}&=&-2(E^r)^2 +2r^4 B^2,\lb{dyonic4.20}\\
F_{\mu\nu}\tilde{F}^{\mu\nu}&=& -4 r^2 E^r B,\lb{dyonic4.21}\\
s&=&\frac12\left([E^r]^2-r^4 B^2\right)+\frac{\kappa^2}2 r^4 (E^r B)^2.\lb{dyonic4.22}
\eea
In view of these results, we see that the nontrivial components of the energy-momentum tensor given in \eq{dyonic4.4} are
\bea
T_{00}&=&Af'(s)\left([E^r]^2+\kappa^2 r^4 [E^r B]^2\right)-Af(s),\lb{dyonic4.23}\\
T_{11}&=&-\frac{f'(s)}A\left([E^r]^2+\kappa^2 r^4 [E^r B]^2\right)+\frac{f(s)}A,\lb{dyonic4.24}\\
T_{22}&=&r^6 f'(s)\left(B^2-\kappa^2 [E^r B]^2\right)+r^2 f(s),\quad T_{33}=\sin^2\theta\, T_{22},\lb{dyonic4.25}
\eea
with the associated trace density
\be
T=2f'(s)\left([E^r]^2-r^4 B^2+2\kappa^2 r^4 [E^r B]^2\right)-4f(s).\lb{dyonic4.26}
\ee
Therefore, inserting \eq{dyonic4.7} and \eq{dyonic4.23}--\eq{dyonic4.26} into \eq{dyonic4.5}, we get the reduced equations
\bea
(r^2 A')'&=&16\pi G r^2\left(r^4 f'(s)(B^2-\kappa^2[E^r B]^2)+f(s)\right),\lb{dyonic4.27}\\
(rA)'&=&1-8\pi G r^2\left(f'(s)([E^r]^2+\kappa^2 r^4 [E^r B]^2)-f(s)\right),\lb{dyonic4.28}
\eea
which are over-determined.
We now show that \eq{dyonic4.27} is contained in \eq{dyonic4.28}  under a certain condition. In fact, inserting the first relation in \eq{dyonic4.15} into the first equation in \eq{dyonic4.16} and using \eq{dyonic4.19}, we have
\be\lb{dyonic4.29}
\left(r^2 f'(s)E^r(1+\kappa^2 r^4 B^2)\right)_r=0.
\ee
Thus, applying \eq{dyonic4.29}, we see that \eq{dyonic4.27} and \eq{dyonic4.28} lead to
\be\lb{dyonic4.30}
(r^2 A')'-r(r A)''=16\pi G r^6 B f'(s)\left(1-\kappa^2 [E^r]^2\right)\left(2B+\frac{rB'}2\right),
\ee
in view of
\be
\frac{\dd s}{\dd r}=(1+\kappa^2 r^4 B^2)E^r (E^r)'+r^4(-1+\kappa^2 [E^r]^2)BB'+2r^3(-1+\kappa^2[E^r]^2)B^2,
\ee
from \eq{dyonic4.22}.
Since consistency requires $(r^2 A')'-r(rA)''=0$, we are led to imposing the equation
\be\lb{dyonic4.31}
B'=-\frac{4B}r,
\ee
in view of \eq{dyonic4.30}. Solving \eq{dyonic4.31}, we find
\be\lb{dyonic4.32}
B=\frac{g}{r^4},
\ee
where $g$ is an integration constant resembling a magnetic charge which may be taken to be positive for convenience.

We remark that the condition \eq{dyonic4.32} is in fact a consequence of the Bianchi identity
\be
\frac1{\sqrt{-g}}\pa_\mu\left(\sqrt{-g}\tilde{F}^{\mu\nu}\right)=0,\lb{dyonic4.33}
\ee
under the radial symmetry assumption, as well. Indeed, inserting \eq{dyonic4.11} and \eq{dyonic4.19} into \eq{dyonic4.33} and observing $\sqrt{-g}=r^2\sin \theta$, we have
\be\lb{dyonic4.34}
(r^4 \sin\theta\, B)_r=0,\quad (E^r)_\theta=0.
\ee
The first equation in \eq{dyonic4.34} indicates that $r^4 B$ is a constant, implying \eq{dyonic4.32}, and the second equation in \eq{dyonic4.34} simply reconfirms that $E^r$ is a radial
function only.

Consequently, it suffices to concentrate on \eq{dyonic4.28}.  In order to understand the physical meaning of the right-hand side of \eq{dyonic4.28}, we recall that
\be
{\cal H}=T^0_0=g^{00}T_{00}=f'(s)\left([E^r]^2+\kappa^2 r^4 [E^r B]^2\right)-f(s),
\ee
in view of the line element \eq{dyonic4.6} and \eq{dyonic4.23}. In other words,  the equation \eq{dyonic4.28} simply reads
\be
(rA)'=1-8\pi G r^2 {\cal H}(r),
\ee
which directly relates the metric factor $A$ to the energy density of the electromagnetic sector through an integration,
\be\lb{dyonic4.37}
A(r)=1-\frac{2GM}r+\frac{8\pi G}r\int_r^\infty {\cal H}(\rho) \rho^2\,\dd\rho,
\ee
where $M$ is an integration constant which may be taken to be positive to represent a mass.

Naturally the energy of the dyonic point charge is given by
\be\lb{dyonic4.38}
E=\int {\cal H}\sqrt{-g}\,\dd r\dd\theta\dd\phi=4\pi \int_0^\infty {\cal H}r^2\,\dd r.
\ee
If this quantity is finite, we may rewrite \eq{dyonic4.37} as
\be\lb{dyonic4.39}
A(r)=1-\frac{2G(M-E)}r-\frac{8\pi G}r\int_0^r {\cal H}(\rho)\,\rho^2\,\dd\rho,
\ee
where, now,
\bea
{\cal H}&=&f'(s)\left(1+\frac{\kappa^2 g^2}{r^4}\right) (E^r)^2-f(s),\lb{dyonic4.40}\\
s&=&\frac12\left((E^r)^2-\frac{g^2}{r^4}\right)+\frac{\kappa^2 g^2}{2r^4}(E^r)^2,\lb{dyonic4.41}
\eea
where $E^r$ is determined by the first equation in \eq{dyonic4.15} or 
\be\lb{dyonic4.42}
D^r=\frac{q}{r^2}=f'\left(\frac12\left[(E^r)^2-\frac{g^2}{r^4}\right]+\frac{\kappa^2 g^2}{2r^4}(E^r)^2\right)\left(1+\frac{\kappa^2 g^2}{r^4}\right) E^r,
\ee
implicitly, which is the radial version of the general equation \eq{dyonicxk48} when gravity is absent. In other words, all the results found earlier to describe a dyonic point charge in the
situation without gravity may directly be carried over to and used in the situation with gravity in our context of a spherically symmetric black hole solution.

We can now make a general description of the associated black hole horizons. First, from \eq{dyonic4.37}, we have $A(\infty)=1$ to fulfill an asymptotic flatness condition. Moreover,
in order to include the classical Reissner--Nordstr\"{o}m solution, we may impose the condition
\be\lb{dyonic4.43}
M<E,
\ee
where the dyon energy $E$ is given by \eq{dyonic4.38}, since the Reissner--Nordstr\"{o}m limit corresponds to $E=\infty$. Hence, in this situation, we have $A(r)\to\infty$ as $r\to0$.  In particular, $A(r)$ has a global infimum. If such an infimum is a minimum
occurring at some $r_0>0$ with $A(r_0)\leq0$, then an event horizon exists: If $A(r_0)=0$, the horizon occurs at $r_0$ and is extremal; if $A(r_0)<0$, there are inner and outer horizons
at some $r_-$ and $r_+$ lying below and above $r_0$, respectively, $0<r_-<r_0<r_+$. Besides, with $A'(r_0)=0$, we see that $r_0$ is determined by the equation
\be\lb{dyonic4.44}
\frac M{4\pi}={\cal H}(r_0) r_0^3+\int_{r_0}^\infty{\cal H}(\rho)\rho^2\,\dd\rho,
\ee
since
\be\lb{dyonic4.45}
A'(r)=\frac{2GM}{r^2}-\frac{8\pi G}{r^2}\int_r^\infty {\cal H}(\rho)\rho^2\,\dd\rho-{8\pi G}{\cal H}(r) r.
\ee
With \eq{dyonic4.37} and \eq{dyonic4.45}, we obtain that the extremal horizon at $r_0$  is given by
\be\lb{dyonic4.46}
\frac1{r_0^2}=8\pi G{\cal H}(r_0).
\ee
Furthermore,  in  nonextremal situations, the outer event horizon at $r_+$ satisfies $A'(r_+)\geq0$ and $A(r_+)=0$, resulting in the Hawking temperature \cite{FK,MW}
\be\lb{dyonic4.47}
T_{\mbox{\small H}}=\frac{A'(r_+)}{4\pi}=\frac1{4\pi r_+}-2G {\cal H}(r_+)r_+,
\ee
where $r_+$ is given by
\be\lb{dyonic4.48}
r_+=2GM-8\pi G\int_{r_+}^\infty{\cal H}(\rho)\rho^2\,\dd\rho,
\ee
implicitly. Alternatively, if
\be\lb{dyonic4.49}
M>E,
\ee
then \eq{dyonic4.39} leads to $A(r)\to-\infty$ as $r\to0$. Hence an event horizon at some $r_0$ where $A(r_0)=0$  is ensured. In addition, to have a unique such horizon
so that the metric factor $A(r)$ changes its sign across the horizon around which the temporal and radial variables flip their roles, it suffices to assume $A'(r)>0$ or
\be\lb{dyonic4.50}
M>4\pi\left(\int_r^{\infty}{\cal H}(\rho)\rho^2\,\dd\rho+{\cal H}(r)r^3\right),
\ee
in view of \eq{dyonic4.45}. 

The common boundary of the conditions \eq{dyonic4.43} and \eq{dyonic4.49} is of course the critical mass-energy condition
\be\lb{dyonic4.51}
M=E,
\ee
under which the expression \eq{dyonic4.39} simplifies itself  into
\be\lb{dyonic4.52}
A(r)=1-\frac{8\pi G}r\int_0^r {\cal H}(\rho)\,\rho^2\,\dd\rho,
\ee
indicating that the possible singularity at $r=0$ is embodied in the local energy term represented by the integral on the right-hand side of the formula.

Finally, it will be informative to obtain the metric factor $A(r)$ asymptotically away from locally regions. For this purpose, note that \eq{dyonic2.18} and \eq{dyonic4.42} imply that
\be\lb{dyonic4.53}
E^r=\frac q{r^2},\quad r\gg1,
\ee
in leading-order approximation. Thus, using \eq{dyonic4.53} in \eq{dyonic4.40} with \eq{dyonic4.41}, we get
\be\lb{dyonic4.54}
{\cal H}(r)=\frac1{2r^4}\left(q^2+g^2\right),\quad r\gg1,
\ee
within the same truncation errors. Therefore, substituting \eq{dyonic4.54} into \eq{dyonic4.37}, we have
\be\lb{dyonic4.55}
A(r)=1-\frac{2GM}r+\frac{4\pi G}{r^2}(q^2+g^2),\quad r\gg1.
\ee
In other words, in leading-order approximation, we {\em universally} recover the classical Reissner--Nordstr\"{o}m solution as should be anticipated. It is interesting that the spherically symmetric asymptotic form \eq{dyonic4.55} is
indifferent to the fine structure of the nonlinear profile function $f(s)$ in the general model defined by \eq{dyonic2.17} and \eq{dyonic2.18}.

Recall that, in \cite{BY}, Brown and York obtained a formulation of a quasilocal mass or energy of a gravitational
system based on the Hamilton--Jacobi formalism of the Einstein theory by Arnowitt, S. Deser, and C. Misner \cite{ADM1959,ADM,ADM1962} so that, in terms of 
the metric \eq{dyonic4.6}, their quasilocal energy contained within the local region stretched to
the `radial coordinate distance' $r>0$ is
 expressed by the simple formula
\be\lb{BY1}
E_{\mbox{\small ql}}(r)=\frac rG\left(1-\sqrt{A(r)}\right),
\ee
and the limit
\be\lb{BY2}
E_{\mbox{\small ql}}(\infty)=\lim_{r\to\infty} E_{\mbox{\small ql}}(r),
\ee
gives rise to the ADM energy or mass \cite{ADM1959,ADM,ADM1962,Carroll,MTW,Wald} of the system in the full space. Thus,  
inserting \eq{dyonic4.39} into \eq{BY1}, we obtain the Brown--York quasilocal mass or energy for a generalized Born--Infeld dyonic black hole metric to be
\be
E_{\mbox{\small ql}}(r)=\frac2{1+\sqrt{A(r)}}\left([M-E]+4\pi\int_0^r {\cal H}(\rho)\rho^2\,\dd\rho\right).
\ee
Furthermore, for $r\gg1$, since $A(r)$ is given by \eq{dyonic4.55} asymptotically, we see that \eq{BY1} renders the result
\be\lb{4.60}
E_{\mbox{\small ql}}(r)=M+\frac{(GM^2-4\pi[q^2+g^2])}{2r}\left(1+\frac{GM}{r}\right)+\mbox{O}(r^{-3}),\quad r\gg1.
\ee
Consequently, the ADM energy or mass of the dyonic black hole coincides with the Schwarzschild mass, $M_{\mbox{\small ADM}}=E_{\mbox{\small ql}}(\infty)=M$, and
is independent of the fine electromagnetic structure of the generalized Born--Infeld model.

\section{Exact dyonically charged black hole solutions with relegated curvature singularities}
\setcounter{equation}{0}

In this section, we consider dyonically charged black holes in the classical Born--Infeld theory defined by \eq{dyonic2.19} and in the exponential model defined by \eq{dyonic3.18}, respectively.  

In the former context, we may insert \eq{dyonic3.15} and \eq{dyonic3.17} directly into \eq{dyonic4.39} to obtain
the exact solution
\bea\lb{dyonic5.1}
A(r)&=&1-\frac{2G(M-E)}r\nn\\
&&-\frac{8\pi G}r\int_0^r \frac{(\beta q^2+\kappa^2 g^2)g^2+(q^2+g^2)\rho^4}{\sqrt{\rho^4+\kappa^2 g^2}(\sqrt{\rho^4+\beta g^2}\sqrt{\rho^4+\beta q^2+\kappa^2 g^2}+\rho^2\sqrt{\rho^4+\kappa^2g^2})}\,\dd\rho\nn\\
&=&1-\frac{2G(M-E)}r-{8\pi G}\left(\frac Q\kappa-\frac{r^2}{3\beta}+\frac{(\kappa^2 Q^2-\beta q^2)}{10\beta\kappa^3 g^2Q}r^4\right)+\mbox{O}(r^6),\quad r\ll1,
\eea
where $\kappa>0$. It is interesting to note that the use of the joint charge $Q$ defined in \eq{dyonic3.17} greatly simplifies the expression. Some properties of this new black hole solution are
discussed as follows.

 Recall that the Kretschmann invariant \cite{Henry,MTW} of the metric \eq{dyonic4.6} reads
\be\lb{dyonic5.2}
K=\frac{(r^2 A'')^2+4(rA')^2+4(A-1)^2}{r^4},
\ee
which indicates that the only possible curvature singularity occurs at the center of matter sources, $r=0$.  Therefore, we are led from \eq{dyonic5.1} and \eq{dyonic5.2} to draw the following conclusions for the
dyonically charged black hole solution obtained.

\begin{enumerate}
\item[(i)] Under the non-critical mass-energy condition
\be\lb{dyonic5.3}
M\neq E,
\ee
the curvature singularity of the dyonically charged black hole at $r=0$ is of the same type as that of the Schwarzschild black-hole solution, $K\sim r^{-6}$.

\item[(ii)] Under the critical mass-energy condition
\be\lb{dyonic5.4}
M=E,
\ee
the metric factor $A(r)$ is analytic at $r=0$ and enjoys the expansion
\be\lb{dyonic5.5}
A(r)=1-{8\pi G}\left(\frac Q\kappa-\frac{r^2}{3\beta}+\frac{(\kappa^2 Q^2-\beta q^2)}{10\beta\kappa^3 g^2Q}r^4\right)+\mbox{O}(r^6),\quad
r\ll1,
\ee
so that the curvature singularity of the dyonically charged black hole at $r=0$ is considerably relegated from that of the Schwarzschild type to the type $K\sim r^{-4}$.

\end{enumerate}

Similarly, we have
\bea\lb{dyonic5.6}
A(r)&=&1-\frac{2G M}r+\frac{8\pi G}r\int_r^\infty \frac{(\beta q^2+\kappa^2 g^2)g^2+(q^2+g^2)\rho^4}{\sqrt{\rho^4+\kappa^2 g^2}(\sqrt{\rho^4+\beta g^2}\sqrt{\rho^4+\beta q^2+\kappa^2 g^2}+\rho^2\sqrt{\rho^4+\kappa^2g^2})}\,\dd\rho  \nn\\
&=&1-\frac{2GM}r+\frac{4\pi G(q^2+g^2)}{r^2}-\frac{\pi G(\beta[q^2-g^2]^2+4\kappa^2 q^2g^2)}{5r^6}+\mbox{O}(r^{-10}),\quad r\gg1,
\eea
in view of \eq{dyonic3.25+}. Note that there is no need now to use the joint charge $Q$ to arrive at a relatively simple asymptotic expression. This result refines the general formula \eq{dyonic4.55} in
the current context of the classical Born--Infeld theory defined by \eq{dyonic2.19}.

We next consider the exponential model \eq{dyonic3.18} whose dyonically charged black hole solutions are studied earlier in \cite{H1,H2,Yang} in the limiting situation where $\kappa=0$.

Since the finite electromagnetic energy of a dyonically charged black hole solution requires $\kappa>0$, we may insert \eq{dyonicH-exp3} into \eq{dyonic4.39} to obtain
\bea\lb{dyonic5.7}
&&A(r)=1-\frac{2G(M-E)}r\nn\\
&&-8\pi G\left(\frac q\kappa+\frac{r^2}{3\beta}-\frac{q r^4}{150\beta\kappa^3 g^2}\left[15\beta+48\kappa^2+15\kappa^2\ln\left(\frac{\beta\kappa^2 g^6}{q^2 r^6}\right)\right]\right)+\mbox{O}(r^{8}),\quad r\ll1,
\eea
in leading order approximation. Thus we can draw similar conclusions regarding the curvature singularity of the metric factor at $r=0$ as those for the Born--Infeld model \eq{dyonic2.19}.

It is interesting to note that the joint charge $Q$ plays the same role as the electric charge $q$, in leading orders, in \eq{dyonic5.1} and \eq{dyonic5.7}, around the center of matter sources of the black hole.

Furthermore,  we may insert \eq{dyonicH-exp2} into \eq{dyonic4.39} to obtain the metric factor $A(r)$ in the region $r\gg1$, which happens to coincide with the asymptotic formula expressed in
\eq{dyonic5.6}, and is a pleasant surprise, indicating that, within the 6th-order approximation, the black hole metric of the Born--Infeld theory \eq{dyonic2.19} and that of the exponential model \eq{dyonic3.18}
are asymptotically non-distinguishable.

Under the critical mass-energy condition \eq{dyonic5.4}, we see from \eq{dyonic5.7} that the curvature singularity at $r=0$ disappears when electricity is switched off, $q=0$. In this situation, the black hole
is purely magnetically charged so that the Hamiltonian \eq{dyonicH-exp} becomes
\be\lb{dyonic5.8}
{\cal H}=\frac1\beta\left(1-\e^{-\frac{\beta g^2}{2r^4}}\right).
\ee
 Observing \eq{dyonic5.4} and inserting \eq{dyonic5.8} into \eq{dyonic4.52}, we find
\be\lb{dyonic5.9}
A(r)=1-\frac{8\pi G r^2}{3\beta}+\frac{8\pi G  h(r)}\beta,
\ee
where
\be
h(r)=\frac1r\int_0^r \rho^2 \e^{-\frac{\beta g^2}{2\rho^4}}\,\dd\rho,
\ee
which is infinitely differentiable at $r=0$ and satisfies $h^{(k)}(0)=0$ for any integer $k\geq0$.  As a consequence, we can substitute \eq{dyonic5.9} into \eq{dyonic5.2} to get
\be\lb{dyonic5.11}
K(r)=\frac{2^9\pi^2 G^2}{3\beta^2} +\mbox{O}(r^{m}),\quad r\ll1,
\ee
where $m$ may be taken to be any positive integer. In particular, this clearly indicates that the curvature singularity at $r=0$ is now removed.

On the other hand, in view of \eq{dyonic5.5}, we emphasize that the curvature singularity for the classical Born--Infeld model \eq{dyonic2.19}
 at $r=0$ can never be completely removed in all situations, unlike what happens in the exponential model \eq{dyonic3.18}.

With \eq{dyonic5.6}, we may compute the Brown--York quasilocal energy for the classical Born--Infeld model  defined by
\eq{dyonic2.19} and that for the exponential model defined by \eq{dyonic3.18} and observe that the contributions of the nonlinear electrodynamics exhibited by the parameters $\beta$ and $\kappa$ arise only in the terms of the order $r^{-5}$ and beyond, for $r\gg1$.  In particular, in leading orders, the quasilocal energy is still as expressed in \eq{4.60}, of course.
 In fact, the terms containing $r^{-5}$ in $E_{\mbox{\small ql}}(r)$ for $r\gg1$ are given by
\bea
&&\frac{21}{16r^5} G^5 M^6-\frac{\pi(q^2+g^2)}{r^5}\left(\frac{35}4 G^4 M^4-15\pi G^3 M^2(q^2+g^2)+4\pi^2 G^2(q^2+g^2)^2\right)\nn\\
&&+\frac\pi{10r^5}\left(\beta[q^2-g^2]^2+4\kappa^2 q^2 g^2\right),
\eea
in which it is seen that $\beta$ and $\kappa^2$ play equal roles in fine-tuning the structure of the quasilocal energy.

\section{Energy conditions}
\setcounter{equation}{0}

From \eq{dyonic4.23}--\eq{dyonic4.25} with \eq{dyonic4.32}, we have
\be\lb{dyonic6.1}
T_0^0=T_1^1=f'(s)\left(1+\frac{\kappa^2 g^2}{r^4}\right) (E^r)^2-f(s),\quad T_2^2=T^3_3=-f'(s)\left(1-\kappa^2(E^r)^2\right)\frac{g^2}{r^4}-f(s),
\ee
which render us with the energy density $\rho={\cal H}=T_0^0$, radial pressure $p_1=p_r=-T^1_1$, and tangential pressure $p_2=p_3=p_\perp=-T^2_2=-T^3_3$, so that
in view of \eq{dyonic6.1} we may express various energy
conditions as follows \cite{D1,D2,HE}.

\begin{enumerate}

\item[(i)] The weak energy condition $\rho\geq0, \rho+p_i\geq0$ ($i=1,2,3$) now reads
\be
{\cal H}=f'(s)\left(1+\frac{\kappa^2 g^2}{r^4}\right) (E^r)^2-f(s)\geq0,\quad
f'(s)\left((E^r)^2+\frac{g^2}{r^4}\right)\geq0.\lb{dyonic6.2}
\ee

\item[(ii)] The dominant energy condition $\rho\geq|p_i|$ ($i=1,2,3$) becomes \eq{dyonic6.2} and $\rho\geq p_i$ ($i=1,2,3$) or
\be\lb{dyonic6.3}
f'(s)\left((E^r)^2-\frac{g^2}{r^4}+\frac{2\kappa^2 g^2}{r^4}(E^r)^2\right)-2f(s)\geq0.
\ee

\item[(iii)] The strong energy condition $\rho+\sum_{i=1}^3 p_i\geq0$ is equivalent to the inequality
\be\lb{dyonic6.4}
f'(s)\left(1-\kappa^2(E^r)^2\right)\frac{g^2}{r^4}+f(s)\geq0.
\ee
\end{enumerate}

It is clear that the classical Born--Infeld model \eq{dyonic2.19} satisfies \eq{dyonic6.2} automatically. 

To check \eq{dyonic6.3} for the Born--Infeld model, we 
 insert \eq{dyonic2.19},
\eq{dyonic2.30}, and \eq{dyonic3.8} into the left-hand side of \eq{dyonic6.3}, to obtain after a lengthy calculation, the result
\bea
&& \frac1{\sqrt{1-2\beta s}}\left((E^r)^2-\frac{g^2}{r^4}+\frac{2\kappa^2 g^2}{r^4}(E^r)^2\right)-\frac2\beta\left(1-\sqrt{1-2\beta s}\right)\equiv\frac{P(r)}{Q(r)},\lb{dyonic6.5}\\
&&P(r)= \frac\beta2(\kappa^2 g^4[\beta q^2+\kappa^2 g^2])^2+2\beta(\kappa^2 g^3)^2(q^2+g^2)(\beta q^2+\kappa^2 g^2) r^4\nn\\
&&\quad\quad\quad+(\kappa g^2)^2(2[\kappa^2 qg]^2+\beta\kappa^2[2q^4+3q^2 g^2+3g^4]+\beta^2 q^2[q^2+g^2])r^8\nn\\
&&\quad\quad\quad+2(\kappa g)^2(2\kappa^2 q^2 g^2+\beta[q^4+g^4])r^{12}+\frac12([2\kappa qg]^2+\beta[q^2-g^2]^2)r^{16},\\
&&Q(r)=R(r) \left(R(r)+\frac12(\beta\kappa qg^2)^2+\frac\beta2([\kappa^2 g^3]^2+2[\kappa g]^2[q^2+g^2] r^4+[q^2+g^2] r^8)\right.\nn\\
&&\quad\quad\quad\left.+(\kappa^2 g^2+r^4)^2 r^4\right),\\
&&R(r)=r^2\sqrt{r^4+\beta g^2}\left(r^4+\kappa^2 g^2\right)^{\frac32}\sqrt{r^4+\beta q^2+\kappa^2 g^2}.\lb{dyonic6.8}
\eea
Hence \eq{dyonic6.5} is positive for $r>0$ and the dominant energy condition holds.

To examine \eq{dyonic6.4}, we  insert \eq{dyonic2.19},
\eq{dyonic2.30}, and \eq{dyonic3.8} into the left-hand side of \eq{dyonic6.4} to get
\bea\lb{dyonic6.9}
&& \frac1{\sqrt{1-2\beta s}}\left(1-\kappa^2[E^r]^2\right)\frac{g^2}{r^4}+\frac1\beta\left(1-\sqrt{1-2\beta s}\right)=\frac1\beta\left(1+\frac{\beta [E^r]^2-1}{\sqrt{1-2\beta s}}\right)\equiv\frac{P_1(r)}{Q_1(r)},\\
&&P_1(r)=\kappa^6 g^8(\beta q^2+\kappa^2 g^2)+g^4(\beta^2 q^4[2\kappa^2-\beta]+\kappa^4 q^2g^2[5\beta-\kappa^2]+\kappa^4[4\kappa^ 2g^4-\beta q^4])r^4\nn\\
&&\quad\quad\quad+\kappa^2 g^4 (6\kappa^2 g^2+q^2[7\beta-\kappa^2])r^8+
g^2([3\beta+\kappa^2]q^2+4\kappa^2 g^2)r^{12}+(q^2+g^2)r^{16},\\
&&Q_1(r)=R_1(r)\left(R_1(r)+r^2[r^8+g^2(\kappa^2 [2r^4+\kappa^2 g^2]+\beta q^2[\kappa^2-\beta])]\right),\\
&&R_1(r)=\sqrt{r^4+\beta g^2}\left(r^4+\kappa^2 g^2\right)^{\frac32}\sqrt{r^4+\beta q^2+\kappa^2 g^2}.
\eea
 These expressions appear complicated. Fortunately, if we contend ourselves with the classical Born--Infeld model \cite{B3,B4} where $\beta=\kappa^2$,  we easily
see that
$P_1(r)>0, Q_1(r)>0$ for $r>0$.
 In other words, the dyonic black hole solution of the classical Born--Infeld theory constructed satisfies the strong
energy condition as well.

For the exponential model \eq{dyonic3.18}, the second inequality in \eq{dyonic6.2} is self-evidently true and the left-hand side of the first inequality satisfies
\be\lb{dyonic6.6}
{{\cal H}}\geq \frac{\e^{-\frac\beta2{\bf B}^2}}\beta \left(W\e^{\frac W2}-\e^{\frac W2}+1\right),\quad W=W(\delta)\geq0.
\ee
It is clear that $W\e^{\frac W2}-\e^{\frac W2}+1\geq0$ for $W\geq0$. As a consequence, the weak energy condition is valid for the exponential model \eq{dyonic3.18}.

To examine \eq{dyonic6.3}, we use \eq{dyonic3.18} to rewrite its left-hand side as
\be
\e^{\beta s}\left(2s+\frac{\kappa^2 g^2}{r^4}(E^r)^2\right)-\frac2\beta\left(\e^{\beta s}-1\right)\geq 2\left(s\,\e^{\beta s}-\frac{\e^{\beta s}}\beta +\frac1\beta\right)\equiv 2h(s).
\ee
It is clear that $h(s)\to\frac1\beta$ as $s\to-\infty$,  $h(s)\to\infty$ as $s\to\infty$, and the only root of $h'(s)=0$ is at $s=0$ where $h(s)$ attains its global minimum, $h(0)=0$.  Hence
$h(s)\geq0$ for all $s$. Therefore, we have shown that the dominant energy condition holds for the exponential model as well.

We then examine the strong energy condition for the exponential model \eq{dyonic3.18} when $\kappa>0$. To do so, inserting \eq{dyonic3.18} into the left-hand side of \eq{dyonic6.4} and applying  \eq{dyonic3.33}, we see
that it renders
\bea
&&\e^{\beta s}\left(1-\kappa^2[E^r]^2\right)\frac{g^2}{r^4}+\frac1\beta\left(\e^{\beta s}-1\right)=\e^{\beta s}\,\left(\frac1\beta+\frac{g^2}{r^4}\left(1-\kappa^2[E^r]^2\right)\right)-\frac1\beta\nn\\
&&=\e^{\beta s}\left(\frac1\beta-\frac{g^2}{r^4}+\frac3{\kappa^2}+\mbox{O}(r^4)\right)-\frac1\beta<0,\quad r\ll1,
\eea
in view of \eq{dyonic3.33}.
Hence \eq{dyonic6.4} fails near $r=0$. On the other hand, for $r\gg1$, we may use \eq{dyonic4.41}  in the left-hand side of \eq{dyonic6.4} to obtain
\be
\e^{\beta s}\left(1-\kappa^2[E^r]^2\right)\frac{g^2}{r^4}+\frac1\beta\left(\e^{\beta s}-1\right)=\frac{(E^r)^2}2+\left(\e^{\beta s}-\frac12\right)\frac{g^2}{r^4}+
\left(\frac12-\e^{\beta s}\right)\frac{\kappa^2 g^2}{r^4}(E^r)^2,
\ee
in leading orders, which is obviously positive. In other words, the left-hand side of  \eq{dyonic6.4} changes sign so that the strong energy condition does not hold for the exponential model \eq{dyonic3.18} when $\kappa>0$.

If $\kappa=0$, a dyonic point charge is of infinite energy. Therefore we should consider the cases of electric and magnetic point charges separately.
In the electric case, $q>0$ and $g=0$, we can use \eq{dyonic3.28} and \eq{dyonic3.29} to determine the electric field explicitly by
\be
\beta (E^r)^2=W\left(\frac{\beta q^2}{r^4}\right).
\ee
Now, with \eq{dyonic3.18} and $s=\frac12 (E^r)^2$, it is clear that \eq{dyonic6.4} holds. In the magnetic case, we
have $q=0$ and $g>0$ so that $s=-\frac{g^2}{2r^4}$ since $E^r=0$ in view of \eq{dyonic3.28}, \eq{dyonic3.29}, and $W(0)=0$.  Hence, with \eq{dyonic3.18}, the left-hand side of \eq{dyonic6.4} assumes the form
\be
h(s)\equiv -2s\,\e^{\beta s}+\frac1\beta(\e^{\beta s}-1),\quad -\infty<s\leq0,
\ee
which satisfies the properties $h(0)=0$, $h(-\infty)=-\frac1\beta$, $h'(s)>0$ for $s<-\frac1{2\beta}$, $h'(s)<0$ for $s>-\frac1{2\beta}$, and $h\left(-\frac1{2\beta}\right)=\frac1\beta(2\e^{-\frac12}-1)>0$. Hence there is a unique quantity $s_0>\frac1{2\beta}$ such that $h(s)<0$ for $-\infty<s<-s_0$ and $h(s)>0$ for $-s_0<s<0$. Consequently we
see that \eq{dyonic6.4} is invalid.  These results explain the
occurrence and  disappearance of the curvature singularity associated with the electric and magnetic point charges, respectively,
in light of the Hawking--Penrose singularity theorems \cite{Haw,HE,HP,MTW,Nab,Pen1,Pen2,Sen,Wald} for which the fulfillment of the strong energy condition is essential.

Consequently, we see that, when $\kappa=0$,  an electrically charged black hole solution fulfills the strong energy condition but a magnetically charged one violates such a condition,
in the exponential model \eq{dyonic3.18}. This
picture explains the reason why the curvature singularity at $r=0$ is eliminated for a magnetically charged black hole solution under the critical mass-energy condition, established earlier,
as demonstrated through \eq{dyonic5.11}, and is consistent with the study of Bronnikov \cite{Bron} that the absence of an electric charge is necessary for
the presence of a charged black hole with a regular center of matter sources.

\section{Conclusions}

In this study,  black hole solutions carrying both electric and magnetic charges,
often referred to as dyonically charged black holes, arising in the generalized Born--Infeld type nonlinear electrodynamics and of finite electromagnetic energies, have been
constructed, which are of the following specific features and characteristics.

\begin{enumerate}

\item[(i)] Being of finite electromagnetic energies, the curvature singularities of these solutions at the center of the matter sources are of the same type as that of the Schwarzschild black hole.

\item[(ii)] Asymptotically, all the solutions approach the Reissner--Nordstr\"{o}m black hole. As a consequence, the Brown--York quasilocal energies of the solutions are asymptotically the same as that of
the Reissner--Nordstr\"{o}m solution. In particular, the ADM mass of such a charged black hole solution is of the Schwarzschild black hole mass, $M$, such that its electromagnetic energy, $E$,
is seen to make no contribution to the
gravitational energy or mass of the system.  

\item[(iii)] The general formulation here provides a convenient framework for constructing dyonically charged black holes in concrete situations, as illustrated with two examples, one
based on the classical Born--Infeld theory and another on an exponential model.

\item[(iv)] For the dyonically charged black hole obtained in the classical Born--Infeld model, the curvature singularity at the center of matter sources, $r=0$,
measured by the Kretschmann invariant, $K$,  is of the same order as that of the Schwarzschild black hole, $K\sim r^{-6}$, when $M\neq E$, and is relegated to
$K\sim r^{-4}$, when $M=E$, which stands out as a critical mass-energy condition. Such a black hole solution satisfies the weak, dominant, and strong energy conditions.

\item[(v)] The dyonically charged black hole obtained in the exponential model enjoys the same local and asymptotic properties as that in the classical Born--Infeld model. However,
when its electric charge is set to zero and the black hole carries a magnetic charge only, its curvature singularity disappears and the black hole becomes regular under the critical
mass-energy condition. Such a black hole satisfies the weak and dominant energy conditions but does not satisfy the strong energy condition.

\item[(vi)] The gravitational metric factors of the dyonically charged black holes in the classical Born--Infeld model and the exponential model are indistinguishable up to the order of $r^{-6}$
for $r\gg1$ but behave rather differently locally with respect to how they depend on electric and magnetic charges,
in addition to the associated total electromagnetic energies. More precisely, the leading-order charge contribution in the former
situation is truly dyonic, of the form $8\pi G \sqrt{\frac{q^2}{\kappa^2}+\frac{g^2}\beta}$, and that in the latter situation is purely electric, of the form $8\pi G \frac q\kappa$, which appears as if the
magnetic charge were turned off  by setting $g=0$ in the former, where $q,g$ are the electric and magnetic charges of the dyonic black hole. Such a picture explains why curvature
singularity of a dyonically charged black hole  in the former situation cannot be removed in the former situation but can be in the latter situation, by imposing the critical mass-energy condition,
$M=E$, and switching off electricity with setting $q=0$, while switching on magnetism with maintaining $g>0$.

\end{enumerate}

\medskip

{\bf Data availability statement}: The data that supports the findings of this study are
available within the article.

\end{document}